\documentclass[11pt]{article}
\usepackage[top=1in, bottom=1in, left=1in, right=1in]{geometry}

\usepackage{framed}
\usepackage{etoolbox}
\AtBeginEnvironment{quote}{\it \color{black}}
\usepackage{subcaption}
\usepackage{amsmath}
\usepackage{subcaption}
\usepackage{mathtools}  
\usepackage{enumitem}
\usepackage[colorlinks=true, citecolor=red]{hyperref}
\usepackage[numbers]{natbib}
\usepackage{url}
\usepackage{booktabs}

\AtBeginEnvironment{quote}{\it \color{black}}
\newcommand{\scl}[2][c]{%
  \begin{tabular}[#1]{@{}c@{}}#2\end{tabular}}

\usepackage{parskip}
\usepackage{mathpazo}

\title{\textbf{Can AI Model the Complexities of Human Moral Decision-Making? A Qualitative Study of Kidney Allocation Decisions}}

\author{\textbf{Vijay Keswani}\\
    {Duke University}
  \and
  \textbf{\quad \quad Vincent Conitzer}\\
    {\quad \quad Carnegie Mellon University}\\
    {\quad \quad University of Oxford}
  \and 
  \textbf{Walter Sinnott-Armstrong}\\
    {Duke University}
  \and
  \textbf{Breanna K. Nguyen}\\
    {Yale University}
  \and
  \textbf{Hoda Heidari}$^*$\\
    {Carnegie Mellon University}
  \and
  \textbf{\quad \quad Jana Schaich Borg}$^*$\\
    {\quad \quad Duke University}
}

\date{}

\begin{document}

\maketitle
\def\thefootnote{*}\footnotetext{Both authors contributed equally to this work.}\def\thefootnote{\arabic{footnote}}

\begin{abstract}
A growing body of work in Ethical AI attempts to capture human moral judgments through simple computational models.
The key question we address in this work is whether such simple AI models capture {the critical} nuances of moral decision-making by focusing on the use case of kidney allocation.
We conducted twenty interviews where participants explained their rationale for their judgments about who should receive a kidney.
We observe participants: (a) value patients' morally-relevant attributes to different degrees; (b) use diverse decision-making processes, citing heuristics to reduce decision complexity; (c) can change their opinions; (d) sometimes lack confidence in their decisions (e.g., due to incomplete information); and (e) express enthusiasm and concern regarding AI assisting humans in kidney allocation decisions.
Based on these findings, we discuss challenges of computationally modeling moral judgments {as a stand-in for human input}, highlight drawbacks of current approaches, and suggest future directions to address these issues.

\end{abstract}

\section{Introduction}

The ever-growing progress in AI technologies has led to its increasing usage in various societal domains to make or assist decisions with a moral valence \cite{bonnefon2016social,gloeckler2022focus,hindocha2022moral, johnston2020preference, sinnott2021ai}. 
Kidney transplant is a prominent example. Kidney allocation decisions have moral implications \cite{chan2024should, sinnott2021ai}, with computational systems currently employed to improve the efficiency of matching kidney donors to patients \cite{freedman2020adapting, roth2004kidney} and for scoring models to quantify the impact of kidney transplants on patients \cite{epts,schwantes2021technology}.
{This role of AI as a moral decision-maker has motivated a growing body of work on developing computational systems that directly capture and represent the relevant human moral values and judgments.} Using tools from preference elicitation literature, 
these studies present participants with several hypothetical moral choice scenarios and learn mathematical models from their responses using standard computational learning apparatus \cite{awad2018moral,bolton2022developing,gloeckler2022focus,johnston2023deploying,kim2018computational,lee2019webuildai,noothigattu2018voting,  siebert2022estimating,srivastava2019mathematical, van2019crowdsourcing}. 
Proponents of this line of work argue for a multitude of benefits of using AI in moral domains, such as (1) the potential for personalized incorporation of moral values \cite{gloeckler2022focus, sinnott2021ai}, (2) gains in efficiency and scale from automation \cite{conitzer2024should, hendrycks2would}, and (3) attenuation of human cognitive limitations \cite{fuchs2023modeling, gandhi2023can, giubilini2018artificial, savulescu2015moral}.
Given these benefits and considering that AI as a moral decision-maker is already becoming a reality, it is critical to evaluate how well AI can model the essential components of human moral decision-making. 

From prior work in psychology, we know human moral decision-making processes are highly complex \cite{capraro2021mathematical, crockett2016computational, crockett2016formal, kleiman2017learning, yu2019modeling}. So it should come as no surprise that AI cannot capture all the nuances involved. 
Indeed recent works highlight a variety of issues associated with AI models trained on judgments for hypothetical moral scenarios; e.g., the difficulty of training reliable AI models when people's moral judgments in these scenarios are noisy and unstable \cite{boerstler2024stability,siebert2022estimating}.
Yet, despite idiosyncrasies (e.g., noisy responses to the same queries), there is still the question of whether an AI can at all capture the \textit{{normative} essence} of human moral decision-making, i.e., how people process morally relevant factors, develop {informed} preferences over moral attributes and values,
and {deliberately} combine the available information to make their final judgment, at least in simple decision-making tasks.
In other words, is AI capable of modeling the {critical components} of human moral decision-making?

 \begin{figure*}
    \centering
    \includegraphics[width=\linewidth]{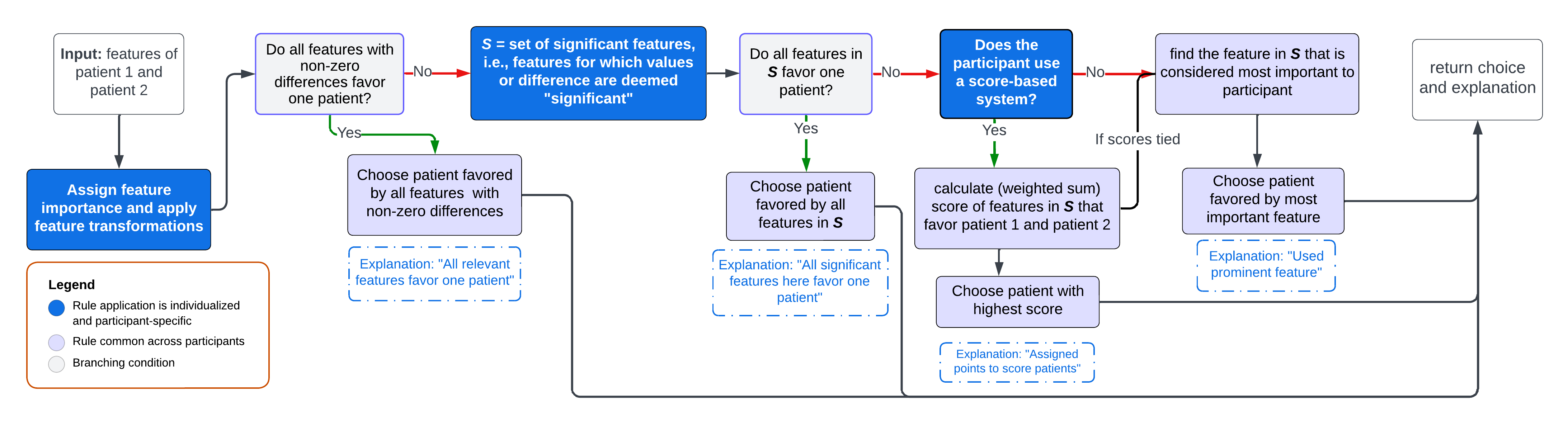}
    \caption{\small A high-level overview of participants' moral decision model for pairwise comparisons in the kidney allocation domain. Note that this only represents participants whose strategies could be clearly represented using decision rules. Not all presented components are used by all participants and this chart doesn't capture cases where participants changed their previously expressed opinions.}
    \label{fig:overall_model_attempt}
\end{figure*}

{
By modeling AI behavior on human moral judgments,
the above-mentioned approaches attempt to employ \textit{human-centered} AI designs \cite{capel2023human}.
For domains like kidney allocation, this translates to computationally eliciting and modeling stakeholders' moral preferences  \cite{freedman2020adapting, yeung2022kidney}.
However, it is unclear whether the assumptions of standard preference elicitation frameworks, such as stable preferences and well-defined preference utility functions, are legitimate for moral domains \cite{keswani2024pros}.    
}
Even if we assume that it is within the scope of current AI frameworks to represent the computational foundations of human moral decision-making,
it is still not obvious what type of AI (e.g., linear models/decision trees or choice of training approach) could do this best.
Employing model classes that do not capture human moral decision-making can reduce AI reliability and accuracy, and subsequently lead to a lack of trust in AI moral decisions.

To better understand if and how an AI can model human moral decision-making, we conducted semi-structured online interviews with 20 participants to discuss their opinions and preferences in the kidney allocation domain.
{We recruited people from the general public as participants for our study, regardless of whether they had experience with kidney allocations or not, to obtain diverse perspectives on how people think kidney allocation decisions ought to be made.}
The discussion focused on 
(a) patient features that participants consider relevant to kidney allocation; (b) their decisions and reasoning on who should get the kidney when presented with pairwise comparisons of patients;
and (c) their attitudes toward using AI in this domain. 
Our contributions center around the following primary findings.
\begin{itemize}
    \item 
    Patient features considered morally relevant to kidney allocation differed across participants in ways that influenced their decision models; e.g., {many (10 of 20) employed feature thresholds or non-linear feature transformations based on when and why they considered the feature relevant.}
    Our analysis sheds light on such differences in feature relevance and underscores the necessity for \emph{individualized} moral decision models.
    \item 
    Participants' reasoning in hypothetical kidney allocation scenarios 
    centered around 
    assigning relative feature importance, pruning the feature set to reduce decision complexity, and/or assigning points to the patients based on favorable attributes (see Figure~\ref{fig:overall_model_attempt}).   
    Yet, decision processes also varied across participants, especially in scenarios perceived to be difficult.
    {Significant individual variation raises key questions about how to come up with a unified AI model that captures the relevant normative considerations of various stakeholders.}
    \item {Some participants (4 of 20)} changed their opinions during the course of the discussions, reflecting a dynamic learning process undertaken when assessing moral scenarios and making moral judgments. 
    \item 
    {Some participants (6 of 20)} expressed uncertainty and lack of confidence in judgments, even noting that they might change their decision based on additional patient information.
    Both variability and uncertainty in decision processes demonstrate the complexity of creating generic methodologies for modeling moral judgments.
    \item {While many participants (13 of 20) expressed AI as potentially useful in this domain, most (17 of 20) advocated for the cautious use of AI and recommended significant human oversight in its implementation.}
\end{itemize}

{
Our work contributes to the evolving HCI literature on the feasibility of human-centered AI designs \cite{auernhammer2020human,capel2023human,lim2019algorithmic,van2022human,yang2020re}, by illustrating the challenges of computationally modeling moral cognition. 
Taking the perspective of AI users and laypeople, our findings call attention to the deliberative and heuristic components of their moral decision-making.
We argue that an appropriate class of AI models should be able to capture these components while setting aside idiosyncrasies such as noise in the application of {well-considered} rules.
Based on our qualitative account, we also find that previously proposed AI moral decision-making models are limited in many ways.
Human moral decision-making is nuanced, non-linear, and dynamic.
In contrast, current AI approaches to capture moral decision-making assume stable preferences, modeled using standard hypothesis classes like linear models \cite{freedman2020adapting,johnston2020preference,kim2018computational,noothigattu2018voting,yeung2022kidney}, which do not accurately capture the dynamic, non-linear nature of how humans make and reason about moral decisions.}
We end with implications for future research in this space, including the necessity of expanded cross-disciplinary exploration of effective elicitation and modeling methods for moral judgments.

\section{Related Work}

\subsection{Moral Cognition and Decision-Making} \label{sec:moral_psych}
Understanding human moral decision-making has been an important part of multiple disciplines, with philosophy \cite{foot1967problem,lim2019algorithmic,singer2005ethics, thomson1984trolley,wolkenstein2018has} and psychology \cite{garrigan2018moral,greene2001fmri,kahane2018beyond,kleiman2017learning,niforatos2020would,yu2019modeling} being the prominent disciplines leading this research. 
The study of moral cognition in these fields broadly involves the exploration of factors that influence people's decisions in moral contexts and their assessment of moral decisions made by others
(see \citet{yu2019modeling} and \citet{malle2021moral} for an overview).
Research on moral cognition has studied judgments for 
real-world \cite{crockett2014harm,feldmanhall2012we} and hypothetical dilemmas \cite{boerstler2024stability,greene2001fmri,haidt2001emotional}, with variations in judgments for different kinds of dilemmas providing concrete insight into the influence of various factors on moral cognition, e.g., intentions \cite{ngo2015two}, emotions \cite{greene2002and}, actions \cite{cushman2013action}, and social norms \cite{pryor2019even}.

Several studies have similarly explored how people perceive the moral aspects of kidney allocation decisions.
\citet{tong2011nephrologists} discuss the perspective of nephrologists who seem to prioritize patient health factors and prefer the societal aspects to be under the purview of policymakers.
\citet{gibbons2017patient} and \citet{tong2012patient} study the preferences of kidney patients and transplant recipients in UK and Australia respectively. They find that patients identify both medical (e.g., donor-patient antigen match) and demographic features (e.g., patient age) as relevant to the decision, with varying views on how to balance transplant utilities and equitable allocation distribution.
\citet{krutli2016fairly} asked people from different backgrounds to evaluate various organ allocation policies (varying equity vs utility impact) and find significant differences in the policies preferred by ethicists, health professionals, and general people. 
All of the studies shed light on the patient features that are morally-relevant to kidney allocation and are helpful to our study design as well.
However, these studies do not explore the processes people undertake to combine patient features when making their decisions and whether these processes can be modeled computationally, which is the primary focus of our study.

In general, moral psychology research on computational models of people's moral decision-making processes is still in the early stages \cite{bello2023computational}.
Recent works have explored computational strategies that can explain moral decision-making, using rule-based approaches \cite{govindarajulu2019quantified}, utility-based frameworks \cite{kleiman2017learning,yu2019modeling}, and reinforcement learning \cite{abel2016reinforcement,hadfield2016cooperative}.
These strategies are often applied to create AI tools for moral domains, as discussed in Section~\ref{sec:moral_ai}.
However, there is limited consensus on which computational approach is most suitable to capture moral cognition \cite{malle2021moral,zoshak2021beyond}.
These strategies also suffer from various limitations, as reported in our findings and other works discussed in Section~\ref{sec:prior_limits}.
Crucially, computational models for moral psychology are primarily employed for their explanatory power--i.e., they are used to explain human moral cognition and the different factors that influence it \cite{crockett2016computational,yu2019modeling}.
Our work pursues a tangential goal of assessing whether AI can serve as an appropriate predictor of moral judgments and \textit{simulate} the processes that humans use to make moral judgments, as has been advertised in the AI literature. 
To that end, our qualitative methodology explores the nuances of human moral decision-making that are currently underrepresented in AI models of moral decision-making.

\subsection{AI for Moral Decision-Making} \label{sec:moral_ai}

In the context of AI, moral decision-making has been predominantly studied to create AI systems
whose decisions reflect the values of relevant stakeholders 
\cite{gabriel2020artificial,huang2024collective,ji2023ai,klingefjord2024human,lee2019webuildai,noothigattu2019teaching,phelps2023models,van2022human}.
\citet{capel2023human} characterize this as {human-centered AI design}, arguing that ``\textit{from a moral standpoint, those who will be impacted should be consulted}'' in the design process.
This design undertakes a participatory approach to derive the relevant ethical principles from people's moral judgments and then trains AI to simulate these judgments when faced with moral dilemmas \cite{birhane2022power,sinnott2021ai,wallach2020machine}.

{
This approach to AI development has been forwarded in several moral domains, 
using standard computational preference elicitation methods to elicit stakeholders' moral preferences \cite{ben2019foundations,chen2004survey,dragone2018constructive}.
For example, in the popular ``Moral Machines Experiment'', \citet{awad2018moral}
present participants with scenarios where an autonomous vehicle faces a choice between causing harm to passengers or pedestrians.
By varying certain attributes across scenarios (e.g., action type or demographics of passengers/pedestrians), they report the factors influencing participants' judgment over preferred actions.
\citet{kim2018computational} and \citet{noothigattu2018voting} further use this ``Moral Machines'' data to train AI models that predict and simulate individual and group moral judgments.
\citet{srivastava2019mathematical} similarly study people's fairness perceptions by analyzing their preferences over various {group fairness} criteria that are commonly used to create \textit{fair} algorithmic prediction systems.
In the domain of resource allocation,
\citet{lee2019webuildai} elicit stakeholder feedback on equity vs efficiency tradeoffs associated with algorithms for food donation transportation services.
\citet{johnston2020preference, johnston2023deploying} similarly obtain people's preferences over resource allocation policies by presenting them with pairwise comparisons of several policies that differ in utility and welfare impacts.
In the healthcare domain, \citet{freedman2020adapting} study participatory methods to align kidney exchange algorithms with stakeholder preferences regarding patient features relevant to such decisions.
\citet{bolton2022developing} similarly explore AI tools for antimicrobial use to balance tradeoffs between immediate patient needs and long-term impacts.
In the electoral domain, \citet{evequoz2022diverse} use computational preference elicitation to create a participatory multi-winner electoral process where voters provide preferences about the representation criteria elected bodies should satisfy and a computational process selects the group of popular candidates who satisfy the voters' representation preferences.}

All of these works on human-centered AI employ standard preference learning strategies.
They present participants with several scenarios/actions/choices of moral valence and ask the participants to make a judgment
(e.g., ``which allocation policy should be preferred'' in \citet{johnston2023deploying} or ``which patient should get the kidney'' in \citet{freedman2020adapting}).
The collected responses are used to train an AI that will attempt to simulate the participant's underlying preference ordering of options (e.g., using Bradley-Terry modeling  \cite{bradley1952rank} or Bayesian learning \cite{kleiman2017learning}).
If the AI is to assist each individual, this process can ensure that the AI's decisions reflect the moral preferences of the individual.
If the AI is to assist several people, then their preferences are appropriately aggregated when encoding them in the AI tool \cite{noothigattu2018voting}.

Note that, by design, all of these AI applications require \textit{accurate} elicitation of human moral preferences.
Given the common mechanisms of preference elicitation across these domains, methodological limitations of these standard elicitation strategies will impact the efficacy of AI moral decision-making in every domain.
Crucially, standard preference elicitation assumes stable preferences, often ordered using linear utility functions \cite{freedman2020adapting, johnston2023deploying, kim2018computational,lee2019webuildai, noothigattu2018voting}.
Our analysis investigates the appropriateness of such assumptions for people's preferences in moral domains.

\subsection{Prior Works on Limitations of Moral Preference Elicitation} \label{sec:prior_limits}

Prior works assessing quantitative models of moral judgments  
have pointed out several limitations of computational preference elicitation, such as instability of the expressed moral preferences \cite{boerstler2024stability,crockett2016computational, helzer2017once, rehren2023stable}, inconsistencies between moral values and judgments \cite{gould2023role,siebert2022estimating}, and overreliance on a narrow set of ethical theories \cite{zoshak2021beyond}.
The inability to characterize the appropriate modeling class for moral decision-making has been associated with potential reliability concerns when training AI to make ethical judgments in real-world settings \cite{aliman2019requisite, cohen2016subjective,hofmann2008comparison,kagan1988additive}.
Utility-based models learned using empirical data also do not perform well for \textit{difficult} moral scenarios in certain studies \cite{boerstler2024stability, kim2018computational}.
Many of these challenges relate to the well-reported complexity of modeling moral decisions (see \cite{capraro2021mathematical,crockett2016formal,ugazio2022neuro}).
Our qualitative analysis also reports issues of moral judgment uncertainty and model class misalignment and provides additional insight into how these modeling errors are related to assumptions of standard preference elicitation.

These prior investigations on limitations of standard moral preference elicitation primarily derive their insights from quantitative and theoretical analyses; for example, by studying moral preference instability using people's responses to the same scenario presented at different times \cite{boerstler2024stability,rehren2023stable} or by modeling errors due to invalid assumptions on moral preferences \cite{kagan1988additive,zoshak2021beyond}. 
Such assessments are concerned with alignment between an AI and humans' moral judgments--i.e., whether the human and the AI would make the same judgment--without worrying about the alignment in the decision processes used to reach those judgments.
Yet, alignment in moral decision processes can be even more crucial.
Limited decision-making transparency is indeed known to reduce trust in AI
\cite{de2020artificial,li2023trustworthy,lima2021human,von2021transparency}.
\citet{lima2021human}, in their study on human perceptions of AI morality, specifically observe that people normatively expect AI tools to justify their judgments and decision processes to the same extent as humans in moral domains.
This requirement of AI alignment in moral decision processes has received little attention in prior works and our work aims to evaluate whether current AI systems for moral domains can achieve this alignment.
Unlike prior works that mainly study decision misalignment, our work focuses on the higher-level misalignment between people's moral decision-making processes and the computational models used to simulate their moral judgments.

\subsection{AI Explainability and Interpretability} \label{sec:xai}
Understanding how an AI model processes information has often been under the purview of explainability and interpretability research.
Using the terminology of \citet{rudin2019stop} and \citet{lipton2018mythos}, 
\textit{interpretability} research characterizes models whose decision-making processes can be understood and evaluated by humans,
and \textit{explainability} research studies how to generate posthoc explanations for AI decisions.

The use of interpretable modeling classes, e.g. linear models or decision trees, has been favored in AI moral designs \cite{vainio2023role,vijayaraghavan2024minimum}, including in several works discussed in Section~\ref{sec:moral_ai}, due to
the potential for computational validation of AI decisions.
In this context, \citet{vijayaraghavan2024minimum} frame interpretability as a useful ``\textit{debugging tool}''.
Explanatory frameworks, on the other hand, are primarily employed to understand the behavior of blackbox AI models or models that are uninterpretable at large scales (e.g., neural networks or random forests) \cite{ali2023explainable}.
These explanatory frameworks can shed light on AI moral judgments, by generating reason-based explanations to determine responsibility for problematic AI moral decisions \cite{baum2022responsibility,mcdermid2021artificial, rane2024enhancing} or by simulating moral reasoning \cite{jin2022make,wei2022chain}.
While interpretability and explainability can help make AI decisions more transparent, this literature is also argued to be limited in building trust in AI moral decisions.
Several works highlight issues like varying interpretability requirements across domains \cite{vijayaraghavan2024minimum}, superficial explanations in high-stakes applications
\cite{lima2022conflict,rudin2019stop}, limitations in discerning responsibility and accountability through algorithmic explanations \cite{lima2022conflict,smith2020no}, and the opacity of algorithmic explanations for non-technical stakeholders \cite{ehsan2021expanding,ehsan2021operationalizing,ehsan2022human}.
Our work discusses similar concerns in the context of models for human moral decision-making in Section~\ref{sec:discussion}.

\section{Methodology} \label{sec:methodology}

{
\subsection{Background on Kidney Allocation}

Kidney allocation policies are generally designed to maximize transplant benefits while reducing racial and socioeconomic disparities \cite{smith2012kidney, waikar2020}.
In the US,
the Organ Procurement and Transplantation Network (OPTN) maintains a national registry for kidney matching and designs allocation policies \cite{transplantliving}.
AI systems are also used to improve the efficiency of kidney matching and allocations
\cite{de2010kidney, schwantes2021technology, waikar2020}.
While these systems and policies are technical in nature, they are still expected to encode relevant ethical values
\cite{hoffmaster2013tragic,ladin2011rational}.
To that end, recent works propose eliciting stakeholder moral preferences to guide the AI and policy in such domains \cite{freedman2020adapting,sinnott2021ai,yeung2022kidney}.
The concrete solutions explored here (and other moral domains--see Section~\ref{sec:moral_ai}) assume stakeholders have stable well-defined moral preferences  \cite{freedman2020adapting, yeung2022kidney}.
Our study investigates these assumptions and questions how well AI can simulate human moral judgments in this domain.
}

\subsection{Recruitment}
Between January and July 2024, we conducted semi-structured online interviews with a total of 20 participants.
All participants were recruited through Prolific and were based in the US. 
Participants were presented with an onboarding survey that contained study details 
and a demographic questionnaire. 
Electronic consent was obtained during onboarding
and a summary of consent information was also presented at the beginning of every interview.

The interviews were conducted virtually and only the audio channel was recorded and transcribed.
Each participant was compensated at least \$20, with proportional bonuses for interviews beyond 45 minutes.
Considering the serious medical nature of the interview topics, participants were also informed about the focus on kidney transplants during onboarding.
They were asked if they or someone close to them had undergone a kidney transplant procedure; if yes, we again mentioned that the study involves discussions on kidney transplant decision-making and participants were asked if they consented to participate in an interview on these topics.
At the beginning of the interview, participants were reminded that ``\textit{you may decline to answer any questions or terminate the interview at any time you choose}''.
The study was approved by an IRB (Institutional Review Board) and a pilot was performed before the full study was launched.

{We recruited people from the general public as participants.
While prior qualitative studies have focused on kidney allocation judgments of medical professionals \cite{tong2011nephrologists} or patients \cite{tong2012patient,gibbons2017patient}, we were interested in how lay people think kidney allocation decisions ought to be made, regardless of whether they had experience with this issue or not.
Our participants varied in age, political orientation, education, and personal experiences with kidney issues, and provided diverse moral perspectives on kidney allocation decisions.
This was particularly beneficial in assessing the complexities and variabilities of moral decision-making across people.
Aggregate participant demographics are presented in Appendix Table~\ref{tab:demographics}.
The number of participants was decided by assessing data quality during collection, following \citet{braun2021saturate}. The data was deemed sufficiently rich to answer our research question after twenty interviews.
We assessed the representativeness of our data with respect to the insights it provided us on people's moral decision-making processes and whether they could be captured using AI models.
We decided to stop recruitment once we had sufficient insights to evaluate differences between people's moral reasoning and computational moral decision models.
}

\begin{figure*}
\centering
\begin{minipage}{.45\textwidth}
  \centering
  \fbox{\includegraphics[width=\linewidth]{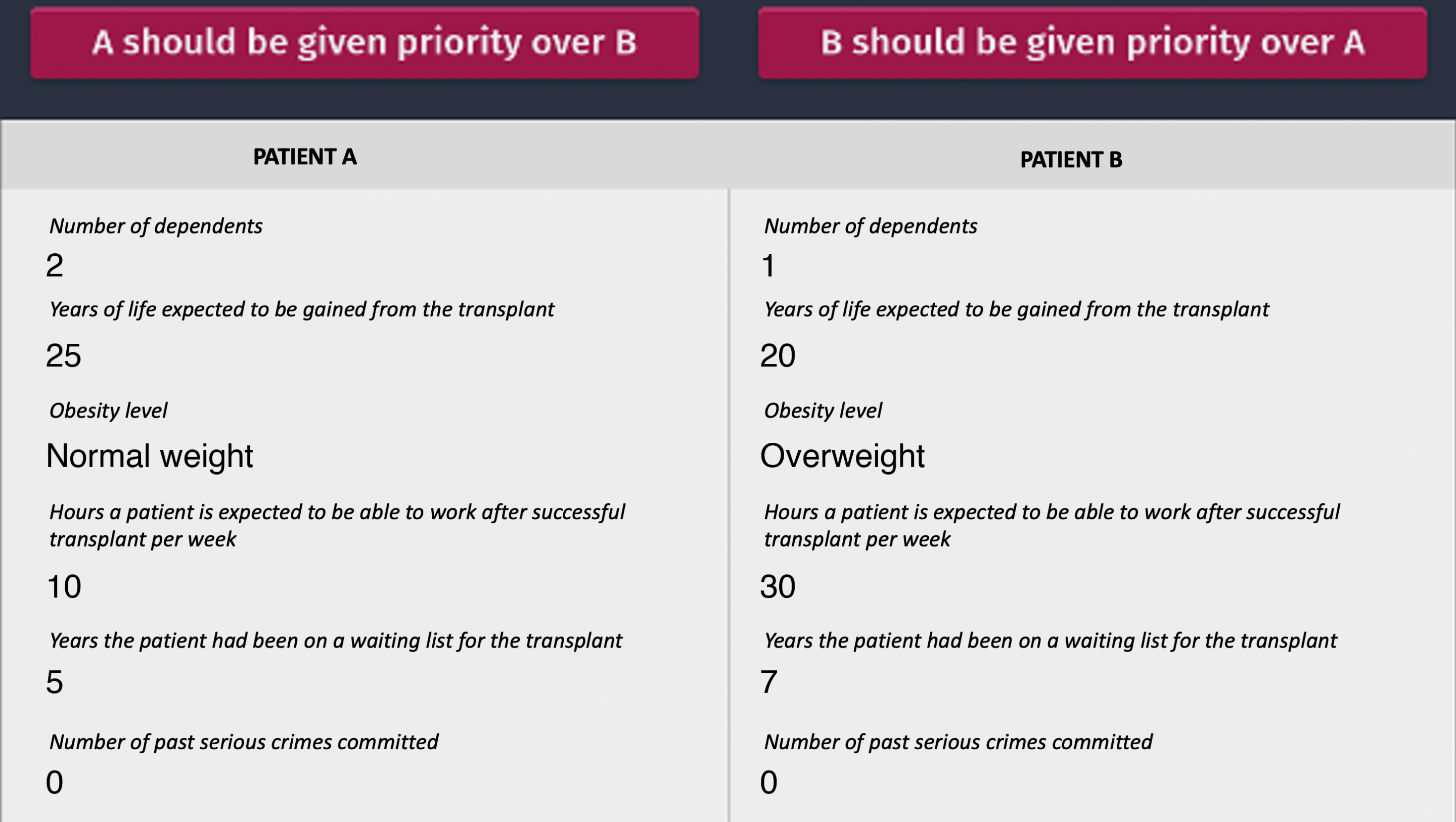}}
  \captionof{figure}{Pairwise Comparison \#2. The \textit{second} pairwise comparison presented to all interview participants.}
  \label{fig:example2}
\end{minipage}\quad\quad\quad\quad
\begin{minipage}{.45\textwidth}
  \centering
  \fbox{\includegraphics[width=\linewidth]{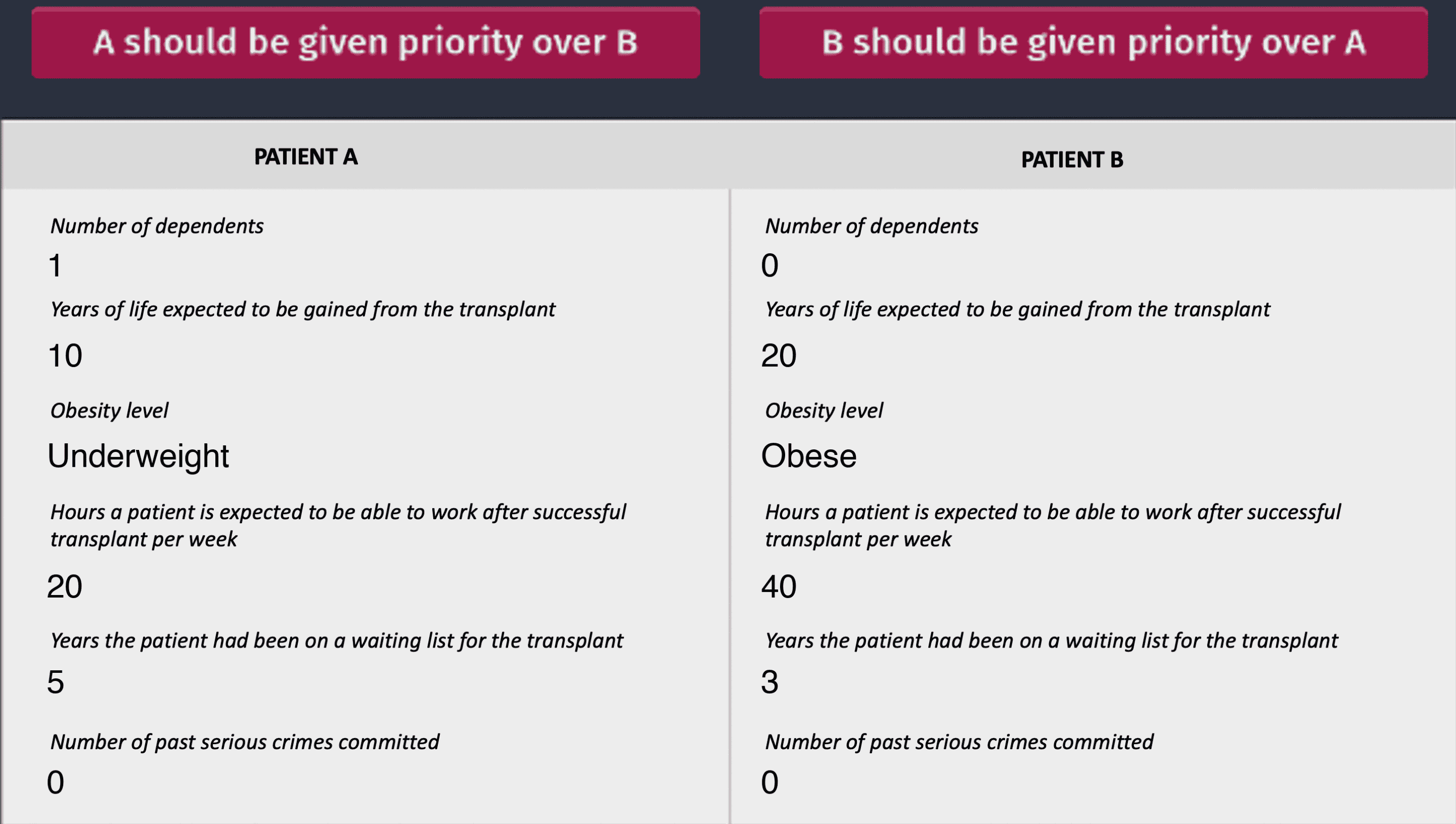}}
  \captionof{figure}{Pairwise Comparison \#3. The \textit{third} pairwise comparison presented to all interview participants.}
  \label{fig:example3}
\end{minipage}
\label{fig:examples}

\end{figure*}

\subsection{Interview Format and Data Collection} \label{sec:interview_format}

The interviews were semi-structured and the high-level script is provided in Appendix~\ref{sec:script}.
All interviews began with a description of the study's purpose and
a short background of the kidney allocation setting,
including information about dialysis procedures for kidney patients
and the various medical and lifestyle restrictions and harms related to dialysis.
This brief overview was meant to emphasize the critical life-and-death nature of kidney allocation decisions.

Following this background, the participants were asked to imagine that they are part of a hospital committee (as a community member) that's designing new rules and regulations for kidney allocation.
They were told that they ``\textit{get to propose the system for kidney allocation the way you believe to be best for reasons you believe to be best}''.
The participant was asked to imagine themself as decision makers while keeping in mind that they're also recommending policies for future kidney allocation decisions.
With this setup, the discussions focused on three aspects of kidney allocation.

\paragraph{(A) Feature relevance.} 
We first asked two open-ended questions: (i) which patient features should be considered when deciding who gets an available kidney, and (ii) which features should not be considered.
Based on their response, participants were presented with additional features (see Appendix~\ref{sec:script} for feature list)
and asked about their relevance as well.
The feature list used is inspired by prior work on morally relevant factors for kidney allocations \cite{freedman2020adapting}.

\paragraph{(B) Decision-making for pairwise comparisons.}
Participants were next presented with 
\textbf{three pairwise comparisons} 
of patient profiles and asked to choose which of the two presented patients in each comparison should be given an available kidney.
Each pairwise comparison describes two patients, A and B, using six features:
(i) years of life expected to be gained from the transplant, (ii) number of dependents, (iii) obesity level, (iv) weekly work hours after transplant, (v) years on the transplant waiting list, and (vi) number of past serious crimes committed.

{We chose these patient features based on past studies on factors that are considered relevant to kidney allocation.
The years of life expected to be gained and years on the transplant waiting list are medical-relevant features that are favored by clinicians, lay people, and policy makers
\cite{freedman2020adapting,krutli2016fairly,yeung2022kidney}.
Past studies on the moral attitudes on kidney allocation have shown that general people also consider non-medical factors like the number of dependents and criminal record to be morally relevant \cite{freedman2020adapting}.
We included obesity due to its medical relevance to transplant outcomes \cite{di2020obesity} and its associations with social and environmental factors \cite{lee2019social}.
Finally, we included the patients' weekly work hours feature since it is often studied as a proxy for patient satisfaction post-transplant \cite{cao2020employment}.
However, in past studies, people have put little to no weight on this feature \cite{boerstler2024stability}; a similar finding in our data would ensure that our work is in line with prior works.   
Note that, this feature set was explicitly constructed to contain a diverse set of medical and social patient attributes so that patient pairwise comparisons pose genuine tradeoffs to participants across different ethical values.
}

{Presenting participants with pairwise comparisons is a popular strategy for AI-based moral preference elicitation \cite{awad2018moral,johnston2023deploying,lee2019webuildai,srivastava2019mathematical}, including in the kidney allocation domain \cite{boerstler2024stability,freedman2020adapting}.
We presented participants with three pairwise comparisons using the above features.
Comparison 1 was personalized for each participant.
Among the features that a participant mentioned to be relevant two were selected.
In this first pairwise comparison, patients A and B differed only across these two features, with one favoring patient A and the other favoring patient B.
For example, if the participant mentioned \textit{years on the waitlist} and \textit{number of dependents} as relevant, then a sample first comparison would be between patient A with 3 dependents and 1 year on the waitlist vs patient B with 0 dependents and 7 years on the waitlist.
All other features will have the same value for patients A and B.
Comparison 1 was constructed this way to present a genuine personalized dilemma to each participant, without overloading them with too many feature differences.
This first comparison was automatically generated during the interview using a simple Python script.

The next two comparisons, presented in Figure~\ref{fig:example2} and \ref{fig:example3}, were kept the same for all participants.
These comparisons were selected from \citet{boerstler2024stability}, who observed 
that many participants changed their judgment when presented with these scenarios at different times.
\citet{boerstler2024stability} demonstrate a significant correlation between the high levels of perceived difficulty for these scenarios and response instability.
Given their findings, these comparisons were well-suited for our setting as they are likely to pose similar moral dilemmas for our participants.
For all comparisons, participants were encouraged to provide their reasoning when making judgments.

}

After these comparisons, five decision strategies were explicitly presented to the participants and
they were asked to rate how well each aligned with their decision-making process.
{These strategies included formal and informal methods studied in the context of moral decision-making.
The formal methods presented the strategies based on a linear model or a decision-tree model, both of which have been proposed for AI moral decision-making \cite{bendel2016annotated,freedman2020adapting, johnston2023deploying}.
The informal methods presented emotion and gut-based strategies studied in moral psychology literature \cite{haidt2001emotional,pletti2016will}--``going with your gut without any explicit procedure'', ``choosing the option you regret the least'', or ``choosing the option that gives you the most satisfaction''.
Certain works in psychology argue for the \textit{dual process} model for moral decision-making, i.e., that moral judgments can be based on emotional or rational processing, depending on the context \cite{greene2001fmri,haidt2001emotional}. 
Following these works, we present both kinds of strategies to participants to obtain self-reports on how well each aligns with their decision-making process.
The complete descriptions of strategies are presented in Appendix~\ref{sec:additional_methodology}.
}

\paragraph{(C) AI attitudes.}
As part of the final questions, participants are asked about their opinions on the use of AI to make decisions in kidney allocation settings. This discussion included their perception of benefits and concerns from AI usage and recommendations on how AI could be safely employed in practice.

\subsection{Analysis and Coding}
The transcripts for the interviews were first automatically generated and then polished manually
to correct transcription errors.
Given our goal of understanding people's moral decision processes through their responses, qualitative analysis was performed using the Reflexive Thematic Analysis (RTA) approach \cite{braun2006using} to identify the relevant themes and patterns in the data.
This approach also provided us with more analytical flexibility than codebook-based approaches.
We followed the approach outlined in \citet{braun2006using} and \citet{byrne2022worked}.
The transcripts were independently coded by two authors, using both semantic codes (descriptive representation of the coded text) and latent codes (interpretations of the coded text), with the latter reflecting the motivating questions for this project. 
Both authors also discussed their codes with each other, and differences in assigned codes were further explored to obtain deeper interpretations of the coded texts.
Higher-level themes were generated by studying commonalities among the codes and developed through multiple discussions of the codes among all the authors.
The themes went through a similar discussion and refinement process, and the primary themes that emerged from this analysis are discussed in detail in the next section.

\subsection{Researcher Positionality}

All authors of this paper are affiliated with academic institutions,
and our backgrounds span various disciplines, ranging from humanities and social sciences to computer and data sciences.
Several authors also have prior experience 
with ethical AI development and have 
conducted qualitative and quantitative studies of people's moral judgments.
These experiences are reflected in our study design, research questions, and analysis methods.
Prior studies of people's moral judgments influenced our interview design.
Our interests center around computationally modeling moral judgments to build ethical AI tools which motivates our primary research questions that focus on obtaining a deeper understanding of participants' judgments and reasoning in moral domains.
Our discussions on the implications of our findings similarly highlight the themes relevant to the complexities associated with people's moral decision-making processes.

\section{Main Findings and Themes} \label{sec:findings}

{
Our findings center around the following observed nuances of participants' moral judgments and reasoning: participants varied in the features they consider morally relevant for kidney allocation (Section~\ref{sec:feature_relevance}) and in their subjective feature valuations
(Section~\ref{sec:valuations}); participants' reasoning for 
pairwise comparisons often involved simple decision rules, with different rules used for different purposes, e.g. 
to prune the feature set or create a feature hierarchy (Section~\ref{sec:decision_processes}); some participants expressed uncertainty in their judgments, limiting the inferences that can be drawn about their moral decision-making process from their responses (Section~\ref{sec:uncertainty}); and most participants were cautiously optimistic about the use of AI for kidney allocation, noting a wide variety of benefits and concerns associated with healthcare AI applications (Section~\ref{sec:ai_attitudes}).
We discuss each of these themes in detail in the subsections below.
}

\subsection{Different Participants Considered Different Features Relevant to Kidney Allocation} \label{sec:feature_relevance}

As mentioned in Section~\ref{sec:interview_format}, participants are first asked about their opinions on relevant/irrelevant patient features for the kidney allocation setting.
For this discussion, we first highlight the extent of variation in the features considered relevant, especially focusing on \textit{how} participants incorporated relevant features in their decision processes.

\begin{itemize}
    \item \textbf{Age and life expectancy.} Almost all participants unpromptedly mentioned age as one of the first features considered relevant to kidney allocation.
    The argument for using age was primarily based on notions that younger patients will benefit more from the kidney (``\textit{lot more life to live}''--P5) 
    or that older patients have a higher chance of adverse medical or surgical outcomes post-transplant.
    Similarly, most participants considered life expectancy from a kidney transplant relevant as well, with many using age as a proxy for this feature (P1, P4, P6, P10, P11, P12, P15, P17).
    Yet, some participants were concerned about age-based discrimination in kidney allocation (P3, P6, P9, P18, P16), arguing that 
    everyone should be ``\textit{treated and kept alive}'' (P9) or expressed preference to allocate based on immediate need (P16).
    To account for this, two participants recommended age-based thresholds that would de-prioritize only those patients who are relatively much older than others (P3, P6).
    Some suggested using these variables cautiously as the expected years of life are not guaranteed (e.g., a patient might meet with an unfortunate unrelated accident) (P3, P20).
    These concerns reflect the tension between \textit{expected} long-term benefits vs immediate needs, a prominent theme further discussed in Section~\ref{sec:valuations}.
    \item \textbf{Number of dependents.} Many participants considered dependents to be a relevant feature since providing the kidney to patients with dependents helps a larger number of people or reduces the potential negative impacts associated with foster care or adoption for child dependents
    (P2, P3, P4, P7, P8, P9, P12, P13, P15, P17, P19, P20). 
    \begin{quote}
    ``\textit{You have to take that step and place a little bit higher value on the life that is supporting other lives}'' (P3)
    \end{quote}
    
    Yet, others did not consider it relevant from a medical perspective or were conflicted about using it as it would imply an unfair treatment of people with no children (P5, P7, P16, P18); e.g.,

    \begin{quote}
    \textit{``just because someone has dependents doesn't mean they're going to take care of the kidney''} (P5)
    \end{quote}
    
    Even among participants who considered it relevant, there were variations in how this feature was accounted.
    Some considered dependents to be a binary feature--zero/non-zero dependents (P3, P9, P18)--and others noted context-dependent relevance
    --e.g. prioritize patients with young dependents
    (P7, P11).
    This variability impacts any downstream process that model dependents similarly for all, e.g. as an independent continuous variable.
    \item \textbf{Time on the waitlist.} Kidney transplant waitlists, managed by centralized institutions, 
    aim to streamline the process of matching kidneys from organ donors to patients who need them.
    While most participants considered the time on the waitlist to be important, they differed in their reasons, which
    included: (a) it is related to the need/urgency of transplant (P3, P5, P9, P18, P19), (b) it denotes the time they have been on dialysis/suffering from the disease (P2, P3, P7, P16, P18, P20) and (c) fairness--those waiting longest should be prioritized (P16, P17, P5, P19). 
    As part of the background information, participants were told that the average life expectancy of a patient on dialysis is 5-10 years and participants with reasoning (b) often used this information to threshold the waitlist time; i.e. considering this feature important only when the value is in 5-10 range
    (P6, P2, P7, P18).
    \begin{quote}
        ``[wait] list is a concern [...] hitting that 5 year mark where you know morbidity becomes an issue.'' (P7)
    \end{quote}

    \item \textbf{Lifestyle choices and past behavior.} 
    Whether and why lifestyle choices were relevant was a contentious issue.
    Some participants brought up lifestyle choices (e.g. if the patient had healthy habits) unpromptedly (P5, P12, P14) while others mentioned it relevant when asked if the past smoking/drinking behavior of the patient matters
    (P2, P3, P4, P7, P8, P9, P10, P19).
    Their consideration also differed in whether it was \textit{forward-looking} or \textit{backward-looking}.
    Forward-looking considerations involved using past behavior to predict future outcomes.
    \begin{quote}
        \textit{``Smoking alcohol and drug use may come into play [...] I think we should probably look for someone who has all of these issues under some level of control before we consider how to allocate kidneys.''} (P7)
    \end{quote}    
    Backward-looking considerations determined patient accountability through their past actions.
    \begin{quote}
        As far as if you have already known about like there was warning that this could happen, then [past behavior] should matter. Like it should absolutely matter.'' (P8)
    \end{quote}
    Others disagreed because these choices are in the past or preferred to check if the patient has undertaken treatment for addiction issues (P11, P16, P18), prioritizing transplant need/future benefit over past actions.
    \item \textbf{Societal factors.} Whether societal factors are relevant was an equally contentious topic, with some participants considering patients' past criminal history (primarily serious violent crimes - P2, P4, P5, P7, P10, P12) or their societal contributions (via job or community impact - P4, P10, P20) to be important.
    Yet, many others believed that societal factors, including criminal record, should not be a factor either because they are not medically relevant (P5, P7), using them would be unfair if they took value in the past (P8, P16, P20), or due to existing systemic biases that could be exacerbated by treatment disparity in kidney allocation settings (P14, P15, P18).
    \item \textbf{Obesity.} The patient's obesity level was considered relevant by some participants who believed it to be related to surgical/medical outcomes (P4, P5, P9, P8, P13, P14, P15, P16, P17, P20) or if associated with responsibility (P16, P19).
    Yet, others either did not consider it relevant (P1) or advised its cautious use in cases where obesity is related to socioeconomic factors (P10) or other underlying medical/genetic conditions (P18).  
    \item \textbf{Work hours post-transplant} - Most participants did not want to consider this feature as it either seemed medically irrelevant or unfairly tied kidney transplants to work productivity.
    Yet some participants thought it could indicate the quality of life post-transplant (P9, P10, P13) and future societal contributions (P10).

\end{itemize}

Beyond these, almost all participants unpromptedly mentioned attributes like race, gender, religion, and sexual orientation, to be irrelevant, citing equity concerns if they are used for kidney allocation.
These findings of relevant, irrelevant, and contentious factors
are consistent with past quantitative work on morally relevant factors in similar domains \cite{chan2024should,freedman2020adapting}.
Yet, note that our qualitative findings go beyond past work and shed light on participants' reasons for feature relevances and the corresponding impact on how these features are modeled in their decision processes.

\subsection{Participants Have Their Own Subjective Preferences and Valuations of Patient Features} \label{sec:valuations}

Implicit to the variability in feature considerations is the fact that participants put differing weights on the subjective values relevant to kidney allocations.
These primary values (listed below) were used to justify the importance assigned to any feature and influenced the tradeoffs experienced by participants when different features favored different patients.

\begin{enumerate}
    \item \textbf{Prioritize patients who gain the most from the kidney}. This preference was reflected in participants' reasons for using features like age and life expectancy, or when determining future medical outcomes using available features, measuring long-term utility from transplant (P2, P3, P4, P5, P9, P8, P11, P14, P17, P18). 
    \item \textbf{Prioritize patients based on urgency.} Some participants valued the urgency or immediate need for a transplant (P1, P2, P4, P7, P9, P10, P16, P18). This was the prominent interpretation of the waitlist feature.
    \item \textbf{Prioritize patients based on the number of people impacted.} When considering  dependents or their societal contributions,
    some participants 
    expressed their priority for patients through whom a larger number of people will positively benefit from the transplant (P2, P3, P9, P8, P12, P19, P20).
    \item \textbf{Prioritize patients expected to act responsibly in the future.} Some \textit{forward-looking} valuations assessed patients' lifestyle choices to judge if they will take care of the transplanted kidney (P2, P3, P4, P5, P10, P13, P15) 
    \item \textbf{De-prioritize patients with past irresponsible conduct.} 
    Alternatively, some had a \textit{backward-looking} consideration of past actions, once again using patients' lifestyle choices, such as actions related to kidney disease, e.g. drinking/smoking habits (P8, P9, P12), or other past misbehavior, e.g., past crimes (P1, P2, P3, P5, P12, P17, P19).
    \item \textbf{Fairness considerations.} Many participants expressed caution about using features that could imply unfair treatment. For features that were considered to be beyond the control of the patients (e.g. obesity due to genetic/socioeconomic factors), some were concerned about potential unfairness in using these features, even if they considered them to be partially relevant to transplant outcomes (P8, P9, P10, P12, P18).
    Other participants also expressed concerns about fairness in using non-medical features (e.g., number of dependents).
    Based on these considerations of fairness, some participants suggested assigning relatively low or no weight to these features or only using them as tie-breakers (P1, P4, P5).
\end{enumerate}

Most participants didn’t use just one valuation but relied on a combination. 
However, conflicts across values (e.g., long-term gains vs urgency) increased the perceived decision difficulty.
Such tradeoffs are methodologically desirable as they elucidate participants' decision models under uncertainty \cite{platt2008risky,tversky1988contingent}, which we discuss in the next section.

Priorities 1--4
align well with a \textit{consequentialist} perspective, allowing for the use of the \textit{expected utility framework} \cite{de2010kidney,harsanyi1977morality,mongin1998expected,schoemaker2013experiments}, and some studies indeed create utility-based computational models for moral judgments based on these valuations \cite{kim2018computational, kleiman2017learning,yu2019modeling}.
Yet
other valuations (5 and 6) do not fit easily within this bucket.
Fairness considerations can manifest as moral constraints for some, 
while for others it can serve as a flexible \textit{utility-regularizing} factor \cite{elalouf2022balancing}. 
With responsibility, forward-looking considerations similarly fit the utilitarian perspective better than backward-looking approaches \cite{chan2024should, wallace1994responsibility}.
While some have argued that these \textit{deontological} considerations can still be folded within a \textit{consequentialist} framework to construct utility functions (see \cite{portmore2022consequentializing}), prior works have also highlighted the difficulty of doing so \cite{black2020absolute, lazar2017deontological}.

\subsection{Variability in Decision Processes and the Use of Decision Rules } \label{sec:decision_processes}

After the feature relevance discussion, participants were presented with three pairwise comparisons between patients A and B. For these comparisons, they decided who should get the kidney and 
were engaged in a discussion on their reasons.
The second and third comparisons are shown in Figures~\ref{fig:example2}, \ref{fig:example3}; 14/20 selected patient A for Comparison~\ref{fig:example2} and 12/20 selected patient A for Comparison~\ref{fig:example3} (i.e., these comparisons were \textit{controversial}).
We discuss the themes that emerged from participants' justifications for their decisions and computational representations of their decision processes.

\subsubsection{\textbf{Participants reported that some feature differences matter and some don't.}} \label{sec:significant_features}
    In all comparisons, patients A and B had different values for at least two features.
    Yet, participants usually 
    reported not taking certain feature differences into account.
    There were three kinds of cases in which feature differences were dismissed.

    \paragraph{(A) Certain features noted irrelevant by the participant.} This was the most straightforward case of participants dismissing features they did not consider relevant at all.
    For example, most participants thought the weekly work hours were irrelevant and dismissed it for Comparison~\ref{fig:example2} and \ref{fig:example3}, both of which had significant differences in this feature.
    \paragraph{(B) Relevant features with small differences were de-prioritized or dismissed.} Certain participants considered some differences in relevant features significant while others were deemed low priority.
    Expectedly, the threshold beyond which a feature difference is significant varied across participants and features. Yet, many participants seemed to use this heuristic to methodically \textit{prune} the set of relevant features in the pairwise comparison
    (P2, P5, P8, P10, P11, P14, P18).
    For instance, P18 made the following justification for not using the years on the waitlist for Comparison~\ref{fig:example2}.

    \begin{quote}
    ``I see the difference of 5 years right between life here expected to be gained [...] And then I do see that, like the years on the transplant list is different, and crimes are zero for both. But so to me, it's only this difference of 2 years about the wait list [...] I think I'm gonna say patient A because the difference between the years being on the waitlist is small. So I think I'll defer to years of life expected.'' (P18) \footnote{Filler words removed from the quotes for ease of presentation.}
    \end{quote}
    Note that, here the importance of the waitlist feature was context-dependent, which we discuss further in Section~\ref{sec:feature_interactions}.

    \paragraph{(C) Certain relevant features were dismissed if their values crossed a particular threshold for both patients.} 
    This use of thresholds to determine feature significance was a recurring phenomenon across participants.
    Beyond thresholds over feature differences,
    participants reported dismissing certain features when the raw value for both patients crossed certain thresholds.
    For example, 
    for certain participants, two patients who both had dependents were at the same standing concerning this feature (P3, P9, P18).
    P18 articulated their reason for this quite well.
    \begin{quote}
        ``I think I'm also not really factoring in one versus two dependents, I think, like the suffering of one child potentially losing a parent versus two children losing a parent [...] I don't know how I feel about weighing sort of that against each other. I think the suffering is enormous in both situations. [...] So the fact that there are dependents, I see maybe as more binary than necessarily thinking about number of dependents.'' (P18)
    \end{quote}
    A similar thought process was expressed in certain cases when determining whether large differences in life expectancy mattered.
    For instance, for Comparison~\ref{fig:example3}, P14 reported choosing patient A because they had a dependent and because 10 years of added life would be greatly beneficial to both the patient and their dependent.

    \begin{quote}
    ``Imagining that, if this person has a child or something, 10 years is going to be the difference between leaving a child and leaving a young adult [...] 
    But I wouldn't sleep well after making this decision.'' (P14) 
    \end{quote}
    The last part of the quote underscores the dilemma faced with this decision.
    Yet, when facing uncertainty, this participant employed a mechanism of reducing the relevant feature set to potentially ease the decision complexity.

    \subsubsection{\textbf{Participants expressed hierarchy over the feature set}}\label{sec:feature_hierarchy}

    By presenting concrete decision scenarios, we were able to elicit the relative importance that participants assigned to various features.
    While the presence of this hierarchy was consistent, the flexibility in following the hierarchy varied across participants.
    In some cases, 
    decisions were based on the most important feature with a significant difference (P4, P11, P14, P16), e.g., as noted by P4 for Comparison~\ref{fig:example3}.

    \begin{quote}
        ``So the years of life, like is the number one priority from all of these factors that I'm looking at. [...] if they're close, then I look at the [...] number of dependents. But that's only if the first factor is close.'' (P4)
    \end{quote}
    The feature hierarchy was relatively more flexible for other participants.
    For instance, P3 reported valuing waitlist time, yet was willing to overrule this feature if someone waiting for a shorter time was in greater need of the kidney.
    \begin{quote}
    ``It's triage [...] If somebody just got put on the list, but they've got one year to live, that has to trump somebody who's been on the list for 10 years, and is of a less need.'' (P3)
    \end{quote}

    \subsubsection{\textbf{Participants assigned context-dependent importance to features.}} \label{sec:feature_interactions}
    The importance assigned to any feature wasn't always independent of the other features.
    Some participants expressed assigning importance to a certain feature $x_1$ based on the value of feature $x_2$ (i.e., \textit{feature interactions}).
    For example, participants P2 and P11 argued that the child dependents of the patient matter only when the dependents are young or when the patient is young (as a proxy of the dependent's age).
    In more complex cases, participants considered a feature in a given pairwise comparison based on other feature values. For instance, when P17 was asked if they are placing more weight on the years on the waitlist feature in Comparison~\ref{fig:example3} relative to Comparison~\ref{fig:example2}, said the following.

    \begin{quote}
    ``I would say it's more in this one, just because, I feel like there's less of a difference in the number of dependents. There's less of a difference in the years expected [...] All those factors are more similar or closer to one another. That's when that's why I place more weight on the amount of years [on the waitlist].'' (P17)
    \end{quote}

    \subsubsection{\textbf{Decision perceived to be quick/easy when all relevant features favored one patient}} \label{sec:easy_decisions}
    Participants seemed to make quick decisions when all important features favored one patient.
    For example, for Comparison~\ref{fig:example2}, P4 decided quickly when all relevant features (number of dependents, life expectancy, and obesity) favored patient A.
    \begin{quote}
        ``Number of hours that they could work isn't relevant to me [...] With how long they've been on the waiting list, not even gonna look at that [...] Patient A has one more dependent than patient B, as well as at least 5 more years expected to be gained. And they have a normal weight, so they don't have any of those like health factors that come into play from obesity or overweight. So I would say patient A.'' (P4)
    \end{quote}

    While participants used multiple relevant features for these decisions, the feature hierarchy wasn't explicit in their reasons, since they all favored one patient (as seen in the above quote).
    Yet, the hierarchy is crucial to understanding their decision process and was expressed more clearly for difficult scenarios with tradeoffs across relevant features.

    \subsubsection{\textbf{Different participants reported different decision processes and rules for difficult decisions}} \label{sec:difficult_decisions}
    The thought process followed when faced with tradeoffs was expectedly more clear for difficult scenarios, where different relevant features favored different patients. 
    The process for difficult scenarios also varied across participants.
    Two recurring processes observed across multiple participants for difficult scenarios are noted below.

    \paragraph{(A) Choosing based on a prominent feature.} 
    Some participants mentioned a single important feature as the reason for their decision (P2, P5, P8, P11, P13, P18, P19, P20).
    For instance, for Comparison~\ref{fig:example3}, P5 decided to use life years gained.
    \begin{quote}
        ``Going through the options, we have a dependent for the person who is underweight but they're only gonna live 10 years [A]. No dependents, but they're gonna live for 20 years [B]. Oh, that it is a real, this one's a really difficult one [...] in a spot of indecisiveness, I went with the amount of years that they would last.'' (P5)
    \end{quote}
    Note that, these participants still employed other features and rules before using the prominent feature.
    In cases like the quote above, they employed their feature hierarchy to decide the prominent feature.
    In other cases, they used the rules discussed in Section~\ref{sec:significant_features} to prune their relevant feature set.
    Heuristics of these kinds eventually led to one single feature.   
    For instance, for Comparison~\ref{fig:example3}, P13 made a different choice than P5 but still decided based on a single feature.
    \begin{quote}
        ``I mean on both ends it seems like the positives and negatives cancel each other out here.  So at the end of the day I would just go with who's been on the waiting list the longest.'' (P13)
    \end{quote}

    \paragraph{(B) Choosing based on points assigned to each patient.} 
    Others found it difficult to 
    solely focus on one feature.
    When faced with difficult scenarios, they instead assigned points or scores to each patient (P1, P3, P4, P15).
    For instance, P15 presented the following reasons for choosing patient A for Comparison~\ref{fig:example2}.

    \begin{quote}
    ``This guy's [A] got 3 pros in terms of the number of dependents, years of life and the normal weight, that gives them 3 points. And then this person [B] has 2 points for the hours the patient expected to work after successful and the years patient have been waiting [...] I weigh dependents to be higher. And I weigh years of life. So patient A.'' (P15)
    \end{quote}    
    Through their reasons, P15 expressed a feature hierarchy but also
    considered all relevant features when deciding.
    Additionally, their last sentence suggests assigning a higher weight to the more important features in this scoring system.
    With a points-based system, there are chances of observing a tie, in which case they could either default to using only the most important feature (as done by P15 for Comparison~\ref{fig:example3}) or tie-breaking features.

\subsubsection{\textbf{Most participants preferred assigning feature weights but couldn't provide precise weights.}} \label{sec:strategies}

After the pairwise comparisons, participants were presented with five decision strategies on the screen and asked how well each aligns with their decision process.
The exact descriptions are noted in Appendix~\ref{sec:strategies_appendix}, but briefly, they can be described as follows: (a) go with your gut, (b) choose the least regretful option, (c) choose the option that gives the most satisfaction, (d) create if-then statements over features to decide, and (e) assigns priority weights to features and add/compare weights to decide.
Appendix Figure~\ref{fig:strategy_likert} presents the aggregated alignment scores for all strategies.
Strategies (d) and (e), were assigned the highest alignment scores by most participants, and many selected strategy (e) as best aligned with their decision process, with some expressing that this strategy is what were trying to do.
Yet, in follow-up discussions on feature weights, these participants only provided relative weights, not precise ones.
\begin{quote}
    ``So I was weighing the chance of success of the transplant more heavily than other factors. So, for instance, obesity levels, the amount of time on the waiting list was a factor. And things that had less direct impact on the success [...] got a lower weight for me, so things like dependents.'' (P7)
\end{quote}

The purpose of this component was to see if participants could self-describe their decision models.
While their responses did provide some insight into this (e.g., their explicit feature hierarchy when discussing weights or if-then examples), in most cases, their statements during this part weren't sufficient to completely characterize their moral reasoning.
In many cases, participants conflated describing their own strategy with the strategy they would recommend as a policy, representing a methodological limitation with this component.
In fact, their expressed reasons for their decisions (as analyzed in the previous sections) seemed to better illuminate their decision processes.

\subsection{{Some Participants Changed Opinions Expressed Previously or Had Unclear Decision Strategies}} \label{sec:uncertainty}

\subsubsection{\textbf{Changing opinions}}
Participants' expressed preferences were not always stable.
Some participants changed their opinion on whether a feature is relevant while reasoning about it (P8, P9, P11, P13).
For these participants, deliberation seemed to bring up other reasons important to them that went against their previous opinions and correspondingly influenced their preferences.
E.g., P8 and P11 both changed their opinion on whether dependents are important when asked for their reasons.
P13 similarly changed their opinion of the weekly work hours feature when deliberating over Comparison~\ref{fig:example2} and employed it as a proxy for quality of life post-transplant.
\begin{quote}
``What's getting me caught up right now is the amount of how much work they would be able to do, even though I said that shouldn't matter. It seems like the quality of life [...] would be better for them [B].'' (P13)
\end{quote}

\subsubsection{\textbf{Inconsistent decision strategies}}
Among participants with clear decision strategies, the point-based and prominent feature system presented two processes for difficult scenarios.
Some other participants, however, presented reasoning that was difficult to model.
For instance, P12 attempted to assign points but struggled to decide which patient they favored when both patients seemed ``equal'' in assigned scores.
In such cases, some participants expressed (P12, P14, P16, P17) needing more information to make a concrete judgment, which they said could potentially change their initial judgment.
Lack of additional information can lead to decision inconsistency if the participant makes different assumptions regarding this information at different times.
In certain other cases, participants did not express a completely clear strategy in their justification or their strategy didn't seem consistent across pairwise comparisons (P7, P8).
This lack of clarity could be a data issue (i.e., that we didn't get enough information about their strategy through discussion) or due to factors that the participant implicitly took into consideration 
but did not explicitly cite during the discussion.

\subsection{Attitudes Toward the Use of AI Ranged From Optimistic to Cautious} \label{sec:ai_attitudes}
Participants were told at the start of the interview that this study aimed to determine if and how an AI could be helpful in kidney allocation settings.
To understand their perspective on AI usage in this domain, at the end of the interview, they were asked whether they think an AI could help assist medical professionals with kidney allocation decisions.

\subsubsection{\textbf{Positive attitudes toward AI}} Most participants thought AI could assist medical professionals. 
When considering the ``basic'' computational abilities of AI tools, some believed that AI could 
filter patient profiles based on pre-determined criteria (P3, P5, P7, P8, P10, P16) or implement decision strategies decided by human experts (P4, P17).
When imagining the potential advanced ``intelligent'' capabilities of AI,
participants thought AI could make objective/impartial decisions (P3, P4, P14, P19, P20) and/or ensure that the decision is free from human biases (P7, P11, P16). 
For some participants, this desire to use an AI reflected their conflicts with using features they believed they were biased toward
(e.g., valuing dependents based on their own parental experiences) and imagined an AI taking a more objective stance 
(P12, P14, P16).
Taking this a step further, participant P3 favored using AI to reduce the emotional burden of making the decision.
\begin{quote}
''Well, I guess part of it is that the computer can do the cold math on it and I can then do what the computer says, and not have to feel regret or guilt about it [...] if I know that a computer has been programmed to weigh criteria in a practical and fair way, then I can trust that computer's answer and not feel guilty.'' (P3)
\end{quote}

\subsubsection{\textbf{Negative attitudes toward AI}}
Most participants also expressed caution about using AI, even when optimistic about their abilities. Their concerns related to 
AI errors 
(P1, P4, P5, P17), AI biases against marginalized groups (P3, P18), and removing the human element from the decision (P7, P8, P10, P18).
Regarding the last point, some participants considered the subjective considerations of kidney allocation decisions to be a positive factor and expressed concern about the potential lack of empathy in AI decisions (P8, P10). 
This is in stark contrast to those who considered AI objectivity to be a positive feature.
Yet, as noted by P10 below, lack of empathy can be scary in life-and-death situations.
\begin{quote}
    ``I think if the AI left the human element, [that] would be the scariest. As soon as it's no longer empathetic and it's no longer `oh person's life has value', that would be the scariest, and easy to do unfortunately.'' (P10)
\end{quote}

Given these cautions,
most participants recommended that AI should not make the final decision; rather even when AI is useful, the final decision should be made by a human (P2, P4, P9, P16, P17, P18).
Many further recommended implementing mechanisms for human oversight over AI functioning and decisions (P3, P4, P5, P7, P12, P18, P20).
Overall, participants' views here reflected the wide spectrum of concerns with AI in healthcare domains.
{Many realized the cognitive flaws of human moral decision-making and were optimistic about AI mitigating these flaws.
Yet, they still expressed belief in the qualifications of human experts and preferred that AI defer the final decision to the experts.
}

\section{Discussion} \label{sec:discussion}

\subsection{Nuances of Human Moral Decision Processes and the Need for Individualized Models}

We undertake a qualitative methodology to 
identify the challenges and nuances associated with the task of computational modeling of moral decision processes.
As mentioned earlier, this enterprise can help build trust in AI's normative decisions by exhibiting that the AI system handles morally relevant factors in a similar manner as humans.
Indeed, decision process misalignment was a concern explicitly expressed by P8 when discussing AI use.
\begin{quote}
    ``My concern, they don't think like a human. They don't think like me, they don't. They don't have my thought process. So what they might think should be ranked as priority I may not think the same way.'' (P8)
\end{quote}
P8's concern is especially validated in light of our main findings on the differences in moral decision processes across participants.
These differences relate to which patient features participants noted as important, their subjective valuations of these features, and decision processes for difficult comparisons.
All of these facets impact how we model moral judgments.
Feature importances and valuations characterize the information used for the judgments
and whether the participant employs any \textit{feature transformation} (e.g. if they threshold the number of dependents or waitlist time).
Their process for difficult decisions determines their choice when faced with tradeoffs across multiple relevant features.
These nuances of moral judgments and the 
{\textbf{observed differences in the moral decision-making process across individuals emphasize the need to simulate people's moral decision processes at an individual level}} using robust modeling/hypothesis classes.

Beyond individualized models, we observed that 
{\textbf{participants undertake a learning process across decisions}.}
An advantage of an interview-based study is that participants were engaged in a discussion that 
allowed them to reason their preferences.
In this setting, participants' decision process seemed to have a \textit{learning} component.
Some changed their preferences as they thought of and expressed additional reasons that went against their previously expressed opinions and some others even mentioned potentially changing their judgment if provided with additional information about the patients (Section~\ref{sec:uncertainty}).
This reflects a dynamic preference construction process which implies that for some participants the moral decision model for two different scenarios can be different.
Prior work has shown that deliberation can generally influence the strength of people's preferences \cite{fournier2011deliberation, oktar2020you, scherer2015trust}.
When eliciting preferences in moral domains, similar discussions/deliberations could be desirable if they lead to better representations of their final preferences \cite{gallo2017heart}.

As we discuss next, these components of moral decision-making impact the performance of computational models that aim to simulate them.
We first demonstrate the drawbacks of current modeling approaches and then highlight the general difficulties associated with computationally capturing the nuances of moral decision-making.

\subsection{Drawbacks of Current Modeling Approaches}

Our findings highlight several drawbacks related to modeling classes employed in prior literature.
For instance, many prior works model moral judgments using \textit{linear utility functions}  \cite{johnston2023deploying, kim2018computational, lee2019social, noothigattu2018voting, van2019crowdsourcing}.
These models assume that any participant's decision process can be parameterized using \textit{independent feature weights} and a score can be computed for each patient by computing the weighted sum of the patient's feature values. Patients with higher scores are prioritized.
For pairwise comparisons, some works even use linear models that only use feature differences without considering individual feature values \cite{johnston2023deploying,kim2018computational}.
Like linear utility models, our participants have a feature hierarchy, assigning relative importance to available features (Section~\ref{sec:feature_hierarchy}), and some indeed use a points aggregation system (Section~\ref{sec:difficult_decisions}).
However, other observations make linear models difficult to justify for moral decisions.
{\textbf{The presence of feature interactions} (Section~\ref{sec:feature_interactions}) \textbf{implies the violation of the feature \textit{independence} assumption}}, since the weight assigned to one feature may not be independent of the other feature values.
{\textbf{The use of thresholds for dismissing certain features based on {significance} demonstrates non-linearity}} and violates the \textit{additivity} assumption of linear models (Section~\ref{sec:significant_features}). 
Dependence on raw values (Section~\ref{sec:significant_features}) further shows that feature differences can be insufficient in characterizing the information used by participants.

Some of these issues, e.g. non-linearity, are mitigated using decision rule-based models. Indeed, 
many participants used {descriptive decision rules} in their justifications (as depicted in Figure~\ref{fig:overall_model_attempt}).
The use of decision rules to model human decision-making has been studied in the behavioral economics literature as well \cite{tversky1972elimination, tversky1974judgment} and
rule-based models have also been proposed for moral domains \cite{bendel2016annotated,gigerenzer2008moral}.
Yet, {\textbf{current decision rule designs by themselves} (including Figure~\ref{fig:overall_model_attempt}) \textbf{are still insufficient for moral decisions}}.
First, we might need an infeasibly large number of rules to handle all possible moral situations.
Second, the rules would also have to be personalized, not just in terms of branching conditions but also individualized feature transformations and rule content, which are not yet accounted for in the proposed rule designs for moral domains.
 
{
Linear and decision rule models have the advantage of being interpretable, which is one reason why they are favored in moral domains \cite{hatherley2024virtues,vijayaraghavan2024minimum}.
However, while desirable, {\textbf{interpretability by itself does not imply alignment with human decision-making}}.
Our interviews show that discussions with participants provide deeper insight into people's moral decision processes.
Yet, discussion-based justifications are not reflected in static interpretable models.
As noted by \citet{watson2022conceptual}, the focus of current interpretability research is on providing static understandings of decisions instead of an ``\textit{iterative exchange}'', which is how people usually justify their decisions to each other in real life.

If the goal is to achieve only accurate prediction given modeling difficulties, then one could argue that neural networks or random forests might be a better fit.
However, employing these models implies sacrificing model interpretability.
While posthoc explanations can be generated for these models using standard tools \cite{lipton2018mythos,suresh2021beyond},
these explanations can also fail to build trust in AI moral decisions for multiple reasons.
First, models like neural networks process information in a different manner than humans; hence, posthoc explanations are unlikely to demonstrate alignment between these models and human moral decision-making, which leads to limited trust in the model's decisions
\cite{franzoni2023black,lima2021human, von2021transparency}.
Second, as argued by \citet{rudin2019stop}, blackbox models are often inappropriate for high-stakes moral domains, especially because current explanatory frameworks still struggle to generate viable and useful explanations.
}

Overall, our findings illustrate the gaps in current approaches related to limited modeling classes.
Yet, one can still ask if there are ways to develop interpretable computational models 
that can better account for the complexities of moral decision-making.

\subsection{Toward An Improved Methodology for Personalized Moral Decision-Aid Tools}

Going beyond modeling choices, our findings demonstrate certain deeper flaws with current elicitation and modeling methodologies used for moral judgments.
Addressing these flaws can potentially lead to more robust AI-based moral decision models by improving the choice of modeling class and elicitation methodologies. 

First, regarding elicitation, prior works eliciting moral judgments in AI-related domains have primarily utilized surveys containing several moral scenarios \cite{awad2018moral, freedman2020adapting, johnston2023deploying, lee2019webuildai, srivastava2019mathematical}.  
In this elicitation design, all that is available in the end is the information presented to the participant about the scenario and their final response.
Once a participant provides several responses, a model can be trained (mainly using the classes discussed above) to simulate a \textit{static/summary} representation of the participant's moral decision process.
However, as we show, {\textbf{participants' process for forming moral preferences is quite dynamic}}, changing opinions and expressing limited confidence in moral judgments in the absence of additional information.
A discussion of moral reasoning provided us with an extra dimension of information during interviews that isn't available via survey designs.
This dimension illuminates participants' subjective feature valuations, the kinds of feature transformations they employ, if and when they change their opinions and preferences, and their deliberations when the presented scenarios are difficult.
Yet, information on these components is not available via survey designs employed in prior works.
An appropriate elicitation and modeling design should capture the deliberation and learning components of human moral decision-making.
Information from these components can help design better models for moral decision-making.
Interview designs can also be improved--visual aids or collaborative boards can better structure discussions of decision processes. 
Given that interviews are costlier than surveys, survey-based methodologies can be complemented with shorter interview/discussion sessions to obtain richer data for each participant.

Second, in terms of modeling, {\textbf{appropriately representing participants' moral decision process requires richer and more interpretable modeling classes than the ones currently employed}}.
For instance, beyond all the issues of linear models discussed earlier, linear models also only consider individual-level variability in feature weights while assuming that the additive process is common to all participants.
Yet, we observe the use of various non-linear feature transformations by many participants, which is beyond the scope of linear models.
Rule-based designs can face similar restrictions if the set of rule choices is not robust enough to 
account for individual-level modeling variability.
Our work also points to certain directions along which better modeling and hypothesis classes can be explored. 
A combination of linear and decision rule-based systems could be better suited to represent participants who relied on a points-based system.
For others who primarily employed decision rules and heuristics, rule-based models would need to be reinforced with other mathematical models to ensure model scalability (e.g., deferring to a linear model when given rules are insufficient in reaching a decision).
Recent work by Rudin et al. \cite{lakkaraju2017learning,rudin2019stop,wang2015falling} on training rule-based systems can be useful for this purpose.
Models that primarily employ decision rules can also ensure better alignment with users and the use of intuitive moral rules across different decision-makers has also been shown to increase trust in decisions \cite{everett2016inference, sacco2017adaptive}.

Inconsistency in participants' moral judgments (as observed in Section~\ref{sec:uncertainty}) can also be problematic for any AI modeling.
Some of the observed participants' {in}consistencies reflect their learning process in the given decision-making domain, i.e., they are inconsistent because they haven't yet formed stable preferences.
{Whether such dynamic preference learning processes should or should not be reflected in an AI is a domain-specific question (indeed attempts to do so have been considered using modern LLMs \cite{jin2022make,wei2022chain}), but that would still require a better understanding of how humans update their moral preferences.
On the technical side, 
\textbf{it is important to investigate the sources of inconsistencies in human moral judgments}.
That way, we can at least attempt to predict when to expect inconsistent moral judgments from stakeholders.
Prior works provides certain fruitful directions for investigations into moral preference instability \cite{boerstler2024stability, helzer2017once,mcelfresh2021indecision} and
elicitation of decision reasoning along with error modeling can also be pursued to study instability.
}

\subsection{Preference Aggregation}

The highlighted variability in participants' moral decision processes also poses challenges to developing aggregated preferences, especially related to mechanisms that appropriately account for the variance in individual-level preferences.
{These challenges are especially pertinent for societal uses of AI, which would include models that are used for kidney allocation in practice.
Aggregated moral preferences should ideally be representative of the preferences of the individuals while encoding appropriate fairness notions to ensure equitable treatment.
However, {\textbf{aggregated preferences cannot represent the underlying population if the elicited individual preference models are inaccurate}}, in the manner we observe.
Modeling issues highlighted in our findings can also impact aggregated group-level preferences.
}

Similar challenges of aggregation have been studied in prior work \cite{el2023strategyproofness,feffer2023moral}.
Yet, aggregation is desirable for policy and AI development as it allows for universality and a uniform application of chosen policies.
To that end, mechanisms like \textit{citizen assemblies}--where a group of selected people engage and deliberate on specific topics to reach a mutual consensus on suitable policies \cite{doherty2023citizen,flanigan2021fair,neblo2010wants}--can be beneficial.
Our qualitative methodology demonstrates that discussion-based elicitation mechanisms provide additional dimensions of information on participants' moral decision processes that are often unavailable through survey-based approaches.
Combining this methodology with the approach undertaken by citizen assemblies could allow for community-based moral decision models that are participatory in their design.

\subsection{Limitations and Future Research Directions}
Our findings inform multiple directions for future research, with several discussed in the paragraphs above.
Additionally, other designs for moral preference elicitation can also be pursued.
{
Prior literature on economic preferences points to decision disparities when choosing between two profiles in a pairwise comparison vs when asked to fill in missing feature values in one profile to make it ``equal'' to the other \cite{erlandsson2020moral,tversky1988contingent}.
Exploring whether similar disparities exist across elicitation methods for AI-related domains can help assess the benefits and drawbacks of different methods.
For instance, one benefit of asking a participant to fill in missing features to make two profiles ``equal'' is that it may provide a clearer picture of their \textit{decision boundary} (i.e. when they are equally likely to choose either profile). 
This method can be useful in AI domains where feature valuations are subjective (e.g., comparing dependents to obesity in kidney allocation, or comparing pedestrian age to action type in autonomous vehicle dilemmas) as it 
can provide better insight into how participants handle tradeoffs across these subjective features.
}

Our study also has certain limitations.
Our participant pool was geographically limited to the US and not representative of all relevant demographic groups.
Moral preferences are socially influenced \cite{graham2016cultural} and cultural differences in moral values relevant to AI applications have been noted in prior works \cite{awad2018moral}. 
Future studies can also explore similar cultural differences in moral decision processes.
Note that our participants took part in an online interview and so might be more receptive to technology, impacting their opinions on AI (Section~\ref{sec:ai_attitudes}). 
Their views may not be representative, but they still highlight certain interesting themes, such as concerns about AI errors/biases and preferences to involve human experts in AI decision pipelines.
While AI opinions are likely to change with improvements in associated technologies, these responses present a valuable snapshot of the multi-faceted perspectives on AI at this current time.
Future work can similarly investigate changes in AI attitudes across time and participant backgrounds.

The presented findings suggest improved model designs, but these will need to be verified using additional empirical evaluations. Due to the interview format, our study sought participant responses for only three pairwise comparisons. Empirical studies on model designs will likely require participant responses to a larger number of moral scenarios.
Our analysis also cannot determine any potential inconsistencies between participants' reported reasoning and implicit considerations that they didn't explicitly express and potential differences in their moral reasoning in experimental vs real-world settings.
Finally, future work can also improve the employed methodology. For example, the five strategies presented to the participants weren't very useful in obtaining self-reports of participants' decision processes;
improved presentation mechanisms can potentially help with enhanced self-reporting of decision strategies.

\section{Conclusion}

Our qualitative study highlights the relevance of understanding the human experience in moral decision-making scenarios.
People's moral judgments for any given scenario aren't made in isolation; they reflect their subjective values, are informed by their backgrounds, and are shaped by the elicitation setup.
A learning and deliberation component makes the decision processes they employ in moral domains more dynamic.
Given these findings, we discuss the drawbacks of current AI approaches to modeling human moral decision-making and present promising directions for future research.
While AI can be helpful (and many participants expressed optimism in its abilities), our study points to the necessity of expanded exploration of effective modeling methods for moral judgments.

\section*{Acknowledgments}
We would like to thank Shivani Kapania for helpful discussions on qualitative methodologies and the anonymous reviewers for their valuable and constructive feedback.
VK, WSA, and JSB are also grateful for the financial support from OpenAI and  Duke.



\bibliography{references}

\clearpage
\appendix

\section{Additional Methodological Details} \label{sec:additional_methodology}

\subsection{Pairwise Comparison Selection}

As mentioned in Section~\ref{sec:methodology}, all participants were presented with three pairwise comparisons.
Their judgments and reasoning for these comparisons shed light on their moral decision processes.

\paragraph{Comparison 1.} The first part of each interview focused on the patient features that participants considered morally relevant.
This included first asking them open-ended questions on which features should and should not be considered and then listing additional features one by one to elicit their opinion on all relevant patient features.
Among the features that a participant mentioned to be relevant, two were selected for the first pairwise comparison.
In this comparison, patients A and B differed across only these two features, with one favoring patient A and the other favoring patient B, with everything else the same.
This first comparison was automatically generated in the background for each participant during the interview using a simple Python script (that took two features as input) that was executed by the interviewer on their own machine.
The comparisons were then presented to the participants using the Zoom screen share feature.

\paragraph{Comparisons 2 and 3.} 
As noted earlier, the next two comparisons are kept the same for all participants (presented in Figure~\ref{fig:example2} and \ref{fig:example3}).
All three comparisons were expected to be difficult for the participants and participants were encouraged to provide their reasoning when making judgments for these comparisons.

\begin{table*}[t]
    \centering
    \footnotesize
    \begin{tabular}{c|c}
        \toprule
        \textbf{Demographic attribute} & \textbf{Statistics}  \\
        \midrule  
        Age  &  Mean: 38.42; Median: 38; Maximum: 53; Minimum: 19; Did not answer: 1\\
        Gender & Male (4); Female (13); Non-binary (2); Did not answer (1)\\
        Race & White (13); Black or African American (6); Did not answer (1) \\
        \scl{Social political orientation} &  \scl{Mean: 2.31 (range: 1 is extremely liberal and 7 is extremely conservative)}\\
        \scl{Economic political orientation} &  Mean: {3.15 (range: 1 is extremely liberal and 7 is extremely conservative)} \\
        Education & \scl{Bachelor or Associate degree (6); Master degree (4), Doctorate (2),\\ Some college but no degree (7), Did not answer (1)}\\        
        Religion & \scl{Christian/Catholic (3); Christian/Non-Catholic (3), Jewish (1), Buddhism (1), Other (12)}\\
        \bottomrule
    \end{tabular}
    \caption{Statistics of participants' demographics self-reported during the onboarding survey.}
    \label{tab:demographics}
\end{table*}

\vspace{0.2in}

\subsection{Decision Strategies Presented to Participants} \label{sec:strategies_appendix}
The following five strategies were discussed with the participants to see how well each aligns with their decision process.
Note that this discussion happened after the decisions and reasons for pairwise comparisons were talked about.

\begin{enumerate}[label=(\alph*)]
    \item ``You go with your gut without any explicit procedure for relating the various features''
    \item ``You think about which option you would regret the least''
    \item ``You choose the option that gives you the most satisfaction and/or pride''
    \item ``You create a bunch of if-then statements related to the patient features that check whether certain features cross particular cutoffs, and string those if-then statements together to decide which patient to prioritize''
    \item ``You give each feature of the patient a priority weight, and then try to add and/or compare the weights for different features to decide which patient to prioritize''

\end{enumerate}

\begin{figure*}
    \centering
    \includegraphics[width=\linewidth]{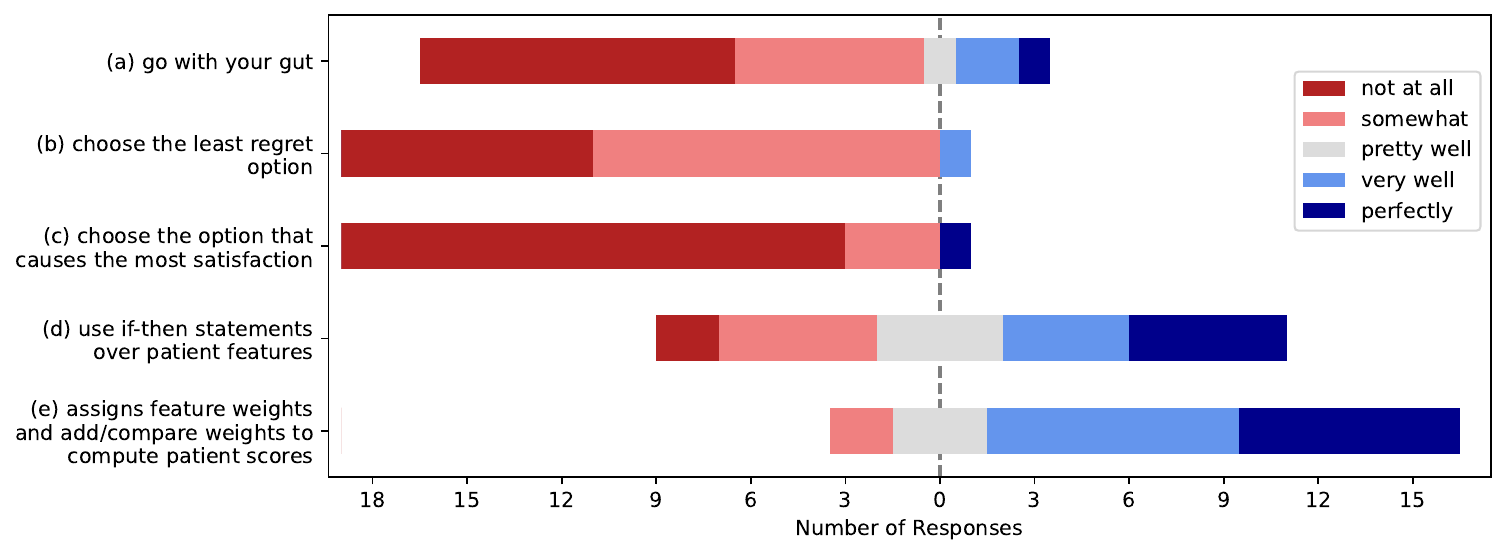}
    \caption{Aggregated alignment scores for all five strategies discussed with the participants.}
    \label{fig:strategy_likert}
\end{figure*}

Participants were asked how well each strategy aligns with their decision process.
For each strategy, they were asked to report the strength of alignment using the following scale: Not at all, Somewhat, Pretty well, Very well, Perfectly.
Based on their responses, Figure~\ref{fig:strategy_likert} presents the alignment scores for all strategies.
Most participants selected the final strategy (e) as the one that best aligns with their decision process, followed by strategy (d).

\clearpage

\section{Interview Script} \label{sec:script}
This section provides the detailed high-level script used for the interviews.
This includes introductory prompts and background provided to each participant.
Also included are all the basic questions posed to the participants.

\vspace{0.1in}
\noindent
{\it
\textbf{Introductory Prompt.} Thank you for speaking with me today. I’m [interviewer-name] with Duke University. Before we begin, let me tell you a little bit about the interview today.
As reflected in the consent form you previously signed, we will record only the audio channel of this interview. Any identifying information you give us will be redacted from the transcripts of the audio recordings. Remember that your participation in this interview is voluntary. You may decline to answer any questions or terminate the interview at any time you choose. Do you have any questions about this?

We are working to better understand how high-stakes decisions that arise within real-world contexts should be made. Our current focus is on trying to understand how a kidney that becomes available should be allocated in kidney transplant centers. What you share with us today will help us determine whether and how an AI might be designed to aid professionals in such decision-making situations. This is ongoing research so we don’t have all the details about how the AI would work. 
But hearing from you will shape how we approach future studies in this area and any decision-support tools we may create. Does this sound ok to you?

Please be open and honest in your answers.  I do have specific questions I want to ask, but we don’t need to stick to a strict question-and-answer format. 
We will begin by discussing how you think organ allocation decisions should be made in ideal circumstances.  Let me give you a little background.

\vspace{0.1in}
\noindent
 \textbf{Conceptions of Morally Desirable Decision-Making.}
The general problem of kidney allocation is that there are not enough donors (live or dead) to supply all patients in need. Roughly 100,000 people in the US alone are in need of kidney transplants. As a result, doctors or hospitals often have to decide which of these patients should receive a kidney when one becomes available.
Kidney transplants can come from either living organ donors or deceased organ donors. For most patients with (chronic) kidney disease, a kidney transplant can help them live longer and have a better quality of life.  While people are waiting for a kidney transplant, they often have to undergo a procedure called dialysis multiple times a week for many hours each visit to help their body remove waste and excess fluid, which is the job a healthy kidney usually does for us. As you can imagine, dialysis is expensive, extremely time-consuming, and makes it difficult for patients to have full-time jobs. Dialysis can also lead to additional medical complications, including infections, sleep problems, bone diseases, low blood pressure, anemia, and depression. Beyond these complications, the average life expectancy on dialysis is 5-10 years making dialysis only a temporary and complicated solution for battling kidney problems. A successful kidney transplant allows patients to get off dialysis and have fewer restrictions on their diet and activities.  On the other hand, kidney patients who don’t get a kidney in time will eventually die from their kidney disease.
Do you have any questions before we continue?

\vspace{0.1in}
\noindent
 \textbf{Decision-making setting.}
 Imagine that your hospital is designing new rules for the allocation of kidneys that become available to be transplanted. The hospital wants to add community members to the committee designing these rules and suppose you are picked to join this committee. You are not bound by any prior rules or regulations. You get to propose the system the way you believe to be best for the reasons you believe to be best.

\vspace{0.1in}
\noindent
 \textbf{Feature relevance.} 
\begin{itemize}
    \item Which features of a patient do you think should be considered when deciding which patient to prioritize for kidney transplant? Why?
    \item Which features of the patient do you believe should NOT be considered when deciding which patient to prioritize for a kidney transplant? Why?
    \item (if they have not mentioned the below features) I am going to list some features of patients that decision-makers could consider when making kidney allocation decisions.  For each feature, please tell me whether you think the feature should, or should not, influence who ultimately gets a kidney when one becomes available? 
    \begin{itemize}
        \item Age
        \item Life years to be gained from the transplant
        \item Number of dependents, such as young children or elderly parents
        \item Past behavior, like drinking or smoking habits, that might have caused their kidney disease. 
        \item Criminal record
        \item Obesity level 
        \item Number of years on the transplant waiting list 
        \item Other health complications, like whether they also have another unrelated disease, like skin cancer
        \item Number of hours they will be able to work per week after a successful transplant 
    \end{itemize}
\end{itemize}

\noindent
 \textbf{Decision scenarios.} 
You told us that features <feature 1> and <feature 2> are important. 
Given that, imagine that Patient A and Patient B are the same in all respects except for these features. 

(Screen share and show a pairwise comparison of two patient profiles A and B so that the above two features conflict and other features are the same for both patients).

Imagine that a kidney becomes available and both of these patients are eligible.  How would you determine which of them should be given the kidney?  Please think out loud and share whatever is going on in your head, even if you aren’t sure how to answer the question yet.

Now I am going to show you more complete profiles of two patients who are on the kidney transplant list. You have to decide which of these two patients should be prioritized to receive a kidney when it becomes available. Like last time, please think out loud about your reasoning and share how you’re making your decision, even if you aren’t sure how you want to answer yet. 

(Comparisons~\ref{fig:example2} and \ref{fig:example3} presented on the screen and discussed.)

\vspace{0.1in}
\noindent
 \textbf{Discussion of decision strategies.} 
Next, I will describe these five different strategies that people use for making decisions for kidney allocations (also presenting all options on the screen).

\begin{itemize}
    \item You go with your gut without any explicit procedure for relating the various features.
    \item You think about which option you would regret the least
    \item You choose the option that gives you most satisfaction and/or pride
    \item You create a bunch of if-then statements related to the patient features that check whether certain features cross particular cutoffs, and string those if-then statements together to decide which patient to prioritize.
    \item You give each feature of the patient a priority weight, and then try to add and/or compare the weights for different features to decide which patient to prioritize

\end{itemize}

First, do you have any questions about these strategies or need any clarification? How well does each strategy describe your decision-making process (using options: Not at all, Somewhat, Pretty well, Very well, Perfectly).

Could you rank the strategies according to how well they describe your decision-making process? Can you tell us how you decided to rank them that way?

\vspace{0.1in}
\noindent 
\textbf{Closing.}
\begin{itemize}
    \item Now that you have seen what the patient comparisons might looks like, do you think an AI would be helpful in any way for these decision-making settings?
    \item Has there been anything surprising or unexpected that has come to mind over the course of thinking about the issues discussed in this interview?
    \item Do you have any feedback for us?

\end{itemize}
 
Thank you so much for taking the time to do this interview and we really appreciate your participation in this study.
}

\end{document}